 \documentclass[nohyper,11pt,letterpaper]{JHEP3}

 \newcommand{\beq}[1]{\begin{equation}\label{#1}}
 \newcommand{\eeq}{\end{equation}}
 \newcommand{\bear}[1]{\begin{eqnarray}\label{#1}}
 \newcommand{\ear}{\end{eqnarray}}

 \catcode`\@=11 \@addtoreset{equation}{section}\catcode`\@=12

 \newcommand{\np}{ {\newpage } }
 \newcommand{\R}{ \mathbb{R} }
 
 \newcommand{\e}{ \mbox{\rm e} }
 \newcommand{\eps}{ \varepsilon }
 \newcommand{\p}{\partial}
 \newcommand{\btd}{\bigtriangledown}
 \newcommand{\btu}{\bigtriangleup}

 \author{V. D. Ivashchuk\\
     Center for Gravitation and Fundamental Metrology,\\
     VNIIMS, 3-1 M. Ulyanovoy Str., Moscow, 119313, Russia
     \\and \\
    Institute of Gravitation and Cosmology,\\
    Peoples' Friendship University of Russia, 6 Miklukho-Maklaya Str.,
    Moscow 117198, Russia \\
     E-mail: \email{ivas@rgs.phys.msu.su}
 }
 \author{D. Singleton\\
     Center for Gravitation and Fundamental Metrology,\\
     VNIIMS, 3-1 M. Ulyanovoy Str., Moscow, 119313, Russia \\
 and \\
 Physics Department, CSU Fresno, \\
 Fresno, CA 93740-8031, USA \\
     E-mail: \email{dougs@csufresno.edu}
 }

 \abstract{ In this paper we consider $(n+1)$-dimensional cosmological model
 with scalar field and  antisymmetric $(p+2)$-form. Using an
 electric composite $Sp$-brane ansatz the field equations for the
 original system reduce to the equations for a Toda-like system
 with $n(n-1)/2$ quadratic constraints on the charge densities. For
 certain odd dimensions ($D = 4m+1 = 5, 9, 13, ...$) and
 $(p+2)$-forms ($p = 2m-1 = 1, 3, 5, ...$) these algebraic
 constraints can be satisfied with the maximal number of charged
 branes ({\it i.e.} all the branes have  non-zero charge densities).
 These solutions  are characterized by self-dual or anti-self-dual charge
 density forms $Q$ (of rank $2m$). For  these algebraic
 solutions with the
 particular $D$, $p$, $Q$ and  non-exceptional  dilatonic
 coupling constant $\lambda$  we obtain general
 cosmological solutions to the field equations and some properties
 of these solutions are highlighted ({\it e.g.} Kasner-like  behavior,
 the existence of attractor solutions). We prove the absence of maximal
 configurations for $p =1$ and even $D$ (e.g. for $D =10$
 supergravity models and those of superstring origin). 
 }

 \title{Composite electric $S$-brane \\
  solutions with maximal number of branes}

\begin{document}

 \section{\bf Introduction}
 \setcounter{equation}{0}

   In this paper we investigate  composite electric
 $S$-brane solutions (space-like analogues of D-branes) in an
 arbitrary number of dimensions $D$ with scalar field and
 $(p+2)$-form. The $(p+2)$-form is considered using a composite
 electric ansatz and the metric is taken as diagonal. All ansatz
 functions for the metric, form field and scalar field are taken to
 depend on only one distinguished coordinate which is taken as
 time-like for the cosmological solutions considered in this paper.
 Previously, related work on cosmological and $S$-brane solutions
 can be found in \cite{BGrIM}-\cite{Ierice} and references therein.

 The procedure we use to investigate our system of
 multi-dimensional gravity plus scalar field plus form field is
 similar to the approach used in \cite{IMC}. This work also studied
 a system with scalar fields and antisymmetric forms field defined on
 the manifold $M_0 \times M_1 \times \dots M_n$ ($M_i$ are Einstein
 spaces and $i \geq 1$). The form fields were taken in the form of a
 composite electro-magnetic p-brane ansatz, the metric was
 block-diagonal, and all scale factors and fields depended upon
 coordinates of $M_0$. Under these conditions the
 original model could be reduced to a gravitating, self-interacting
 sigma-model on $M_0$ with quadratic ``constraints'' on the charge
 densities. These constraints came from the non-diagonal part of
 the Einstein-Hilbert equations. It was shown that the constraints
 could be satisfied for certain ``non-dangerous'' intersection rules
 of the branes \cite{IMC}.  In the present work we use the same
 sigma-model approach to show that it is possible to satisfy the
 constraints maximally ({\it i.e.} all the branes carry non-zero
 charge densities) in certain odd dimensions. We then examine
 cosmological solutions in these odd dimensional cases, and discuss
 some of their interesting features such as their Kasner-like
 behavior.

 The  importance of studying  solutions
 with ``maximal'' number of branes is related
 to research of oscillating behavior of
 cosmological solutions near the singularity
 \cite{IMb1,DH,DHN}, e.g. using the so-called
 billiard approach \cite{IMb1}.  In \cite{DH}
 and other related works (for a review, see \cite{DHN}) it
 was argued that in superstring cosmology one gets chaotic behavior
 in terms of the ``oscillations'' of Kasner parameters as one
 approaches the cosmological singularity. In these works the
 maximal number of electric branes were considered.

 In the next two sections we will give the set up for the system
 of $D=n+1$ dimensional gravity, plus a scalar
 field, plus a $(p+2)$-form field. For the conditions considered in
 this paper (diagonal metric, composite $Sp$-brane ansatz for the
 antisymmetric $(p+2)$-form field, and all the ansatz functions
 depending only on one coordinate) this complex system can be
 reduced to a 1-dimensional $\sigma$-model. This transformation
 greatly helps in studying the solutions of the system.

 In section   4 we consider the quadratic constraints for the charge
 densities of the branes. We find that these constraints have ``maximal''
 solutions with all non-zero brane charge densities  in particular odd
 dimensions with particular form fields: $D= 5, 9, 13, \dots$ and $p =
 1,3,5 \dots$, respectively. We prove also the absence of maximal
 configurations for $p =1$ and even $D$.

 In  section 5 we investigate
 cosmological solutions to the field equations for these odd
 dimensions.  We look at the  proper time behavior of the
 simplest of these solutions. We also show that certain solutions
 exhibit Kasner-like behavior at these early times.

 \section{\bf D-dimensional gravity coupled to $q$-form and scalar field}
 \setcounter{equation}{0}

 Here as in \cite{IMC} we consider the model governed by the action
   \beq{2.1i}
    S =
       \int_{M} d^{D}z \sqrt{|g|} \left[ {R}[g] -
       g^{MN} \partial_{M} \varphi \partial_{N} \varphi
    -  \frac{1}{q!} \exp( 2 \lambda \varphi ) F^2 \right],
   \eeq
 where $g = g_{MN} dz^{M} \otimes dz^{N}$ is the metric,
 $\varphi$   is a  scalar field, $\lambda \in  \R$ is a
 constant dilatonic coupling and
   \beq{2.2i}
   F =  dA =
   \frac{1}{q!} F_{M_1 \ldots M_{q}}
   dz^{M_1} \wedge \ldots \wedge dz^{M_{q}} ,
   \eeq
 is a $q$-form, $q =  p +2 \geq 1$, on a $D$-dimensional manifold
 $M$.

 In (\ref{2.1i}) we denote $|g| = |\det (g_{MN})|$, and
   \beq{2.3i}
   F^2 =
         F_{M_1 \ldots M_{q}} F_{N_1 \ldots N_{q}}
         g^{M_1 N_1} \ldots g^{M_{q} N_{q}},
   \eeq

 The equations of motion corresponding to  (\ref{2.1i}) are
   \bear{2.4i}
   R_{MN} - \frac{1}{2} g_{MN} R  =   T_{MN},
   \\
   \label{2.5i}
   {\btu}[g] \varphi -  \frac{\lambda}{q!}
    e^{2 \lambda \varphi} F^2 = 0,
   \\
   \label{2.6i}
   \nabla_{M_1}[g] (e^{2 \lambda \varphi}
    F^{M_1 \ldots M_{q}})  =  0.
   \ear
 In (\ref{2.5i}) and (\ref{2.6i}), ${\btu}[g]$ and ${\btd}[g]$ are
 Laplace-Beltrami and covariant derivative operators corresponding
 to  $g$. Equations (\ref{2.4i}), (\ref{2.5i}) and (\ref{2.6i})
 are,  respectively, the multidimensional Einstein-Hilbert
 equations, the ``Klein-Fock-Gordon" equation for the scalar field
 and the ``Maxwell" equations for the $q$-form.

 The source terms in (\ref{2.4i}) can be split up as
   \bear{2.7i}
   T_{MN} =   T_{MN}[\varphi,g]
   + e^{2 \lambda \varphi} T_{MN}[F,g],
   \ear
   with
   \bear{2.8i}
   T_{MN}[\varphi,g] =
   \p_{M} \varphi \p_{N} \varphi -
   \frac{1}{2} g_{MN} \p_{P} \varphi \p^{P} \varphi,
   \\
   T_{MN}[F,g] = \frac{1}{q!} \left[ - \frac{1}{2} g_{MN} F^2
   + q  F_{M M_2 \ldots M_{q}} F_{N}^{~ M_2 \ldots M_q}\right] ,
   \label{2.9i}
   \ear
 being the stress-energy tensor of the scalar and $q$-form,
 respectively.

 Let us consider the manifold
   \beq{2.10g}
    M = (u_{-}, u_{+})  \times \R^{n}
   \eeq
 with the metric taken to be diagonal and of the form
   \beq{2.11g}
    g= w \e^{2{\gamma}(u)} du \otimes du +
    \sum_{i= 1}^{n} \e^{2 \phi^i(u)} \eps_i dy^i \otimes dy^i ,
   \eeq
 where $w=\pm 1$, and $u$ is a distinguished coordinate. The metric
 ansatz functions $\gamma (u), \phi^i (u)$, the scalar field
 $\varphi (u)$ and the $q$-forms are assumed to depend only on
 $u$. For definiteness one can think of $u$ as the ``time"
 coordinate but when $w=+1$, $u$ is space-like. Here
   \beq{2.14g}
    \eps_i = \pm 1
   \eeq
 are signature parameters with  $i= 1, \ldots, n$.
 When $u$  is  time-like  and all $\eps_i =  1$
 the solutions are cosmological.   The functions
 $\gamma,\phi^i$: $(u_-,u_+) \to \R$ are smooth.

 In order to deal in a general way with the different possible
 indices for the various forms ($q$-form, volume form) we define
   \beq{2.15g}
   \Omega_0 = \{ \emptyset, \{ 1 \}, \ldots, \{ n \},
              \{ 1,2 \}, \ldots, \{ 1,  \ldots, n \} \}
   \eeq
 which is the set of all subsets of
   \beq{2.25n}
   I_0 \equiv\{ 1, \ldots, n \}.
   \eeq
 These sets indicate the number and ranges of the indices of the
 $q$-forms.

 For any $I = \{ i_1, \ldots, i_k \} \in \Omega_0$ with
 $i_1 < \ldots <
 i_k$, we define a form of rank $d(I) \equiv  k$
   \beq{2.17i}
   \tau(I) \equiv dy^{i_1}  \wedge \ldots \wedge dy^{i_k},
   \eeq
 The corresponding brane submanifold is described by coordinates
 $y^{i_1},   \ldots,  y^{i_k}$. We also define the ${\cal E}$-symbol as
    \beq{2.19e}
          {\cal E}(I) \equiv  \eps_{i_1} \ldots \eps_{i_k}.
    \eeq

 We adopt the following electric composite $Sp$-brane ansatz for
 the field of the $(p+2)$-form
   \beq{2.27n}
       F = \sum_{I \in \Omega_{e}} d \Phi^{I} \wedge \tau(I)
   \eeq
    where the set
   \beq{2.d1}
    \Omega_{e} \equiv \{I \in \Omega_{0}|  d(I) = q - 1 = p + 1\}
   \eeq
 contains all subsets of $\Omega_0$ of the ``length'' $p+1$, {\it
 i.e.} of the form $\{ i_0, i_1, ..., i_p \}$.

 We assume that the scalar potential and the scalar field only
 depend on the distinguished coordinate
    \beq{2.28nn}
       \Phi^I  = \Phi^I(u) ~, ~~  \varphi = \varphi(u).
    \eeq

    \section{\bf $\sigma$-model representation with constraints}
    \setcounter{equation}{0}

    \subsection{\bf $\sigma$-model}

 The system of the previous section can be greatly
 simplified. In \cite{IMC} (see Proposition 2 in \cite{IMC}) it was
 shown that the diagonal part of Einstein  equations (\ref{2.4i})
 and the equations of motion (\ref{2.5i})--(\ref{2.6i}), for the
 ansatz given in (\ref{2.11g}), (\ref{2.27n})--(\ref{2.28nn}), are
 equivalent to the equations of motion for a 1-dimensional
 $\sigma$-model with the action (see also \cite{GrIM,IMJ})
   \beq{2.25gn}
    S_{\sigma} =
     \frac{1}{2} \int du {\cal N} \left[ G_{ij} \dot \phi^i \dot \phi^j
     + \dot \varphi^{2}   + \sum_{I \in \Omega_{e}}
     {\cal E}(I) \exp[-2U^I(\phi,\varphi)](\dot\Phi^I)^2
   \right],
   \eeq
 the overdots represent differentiation with respect to the
 distinguished coordinate, {\it i.e.} $\frac{d}{du}$.

 The factor ${\cal N}$ is the lapse function given by
   \beq{2.24gn1}
     {\cal N}= \exp(\gamma_0-\gamma)>0
   \eeq
 with the definition
     \beq{2.24gn}
     \gamma_0(\phi)   \equiv \sum_{i=1}^n \phi^i,
    \eeq
 Next the factor in the exponent is given by
     \beq{2.u}
       U^I = U^I(\phi,\varphi)= - \lambda \varphi +
                               \sum_{i \in I} \phi^i.
     \eeq
 Finally,
   \beq{2.c}
    G_{ij}= \delta_{ij}- 1
   \eeq
 are components of the ``pure cosmological'' minisupermetric matrix,
 $i,j= 1, \dots, n$  \cite{IM2,IMZ}.

 In this rewriting of the system the generalized ``Maxwell
 equations'' of (\ref{2.6i}) become
   \beq{5.29n}
    \frac d{du}\left(\exp(-2U^I) \dot \Phi ^I \right)=0.
   \eeq
 They can be readily  integrated to give
   \beq{5.29na}
    \dot \Phi ^I= Q(I) \exp(2U^I),
   \eeq
 where $Q(I)$ are constant charge densities and $I \in \Omega_{e}$.

 We will analyze the $\sigma$-model representation of the original
 system in the harmonic gauge where
   \beq{4.1n}
        \gamma= \gamma_0,  \quad  {\cal N} = 1.
   \eeq

 We now further simplify the $\sigma$-model in (\ref{2.25gn}) by
 introducing collective variables $x=(x^A)=(\phi^i,\varphi)$ and a
 ``truncated'' target space metric
    \bear{2.35n}
     \bar G=\bar G_{AB}dx^A\otimes dx^B=
     G_{ij}d\phi^i\otimes d\phi^j+
     d\varphi \otimes d\varphi, \\ \label{2.36n}
     (\bar G_{AB})=\left(\begin{array}{cc}
     G_{ij}&0\\
     0& 1
    \end{array}\right).
    \ear
 $U^I (x)$ is defined in (\ref{2.u}). It can be written as $U^I(x)=U_A^I
 x^A$
    \beq{2.38n}
      (U_A^I)=(\delta_{iI},- \lambda)~,
   \eeq
 where
    \beq{2.39n}
    \delta_{iI}\equiv \sum_{j\in I} \delta_{ij}=
    \begin{array}{ll}
                     1, &i \in I; \\
                     0, &i \notin I;
    \end{array}
    \eeq
 is an indicator of $i$ belonging to $I$. For fixed charge
 densities $Q(I)$, $I \in \Omega_{e}$, the equations of motion for
 the $\sigma$-model  in (\ref{2.25gn}) are now equivalent to the
 equations from the Lagrangian
   \beq{5.31n}
     L_Q=\frac12 \bar G_{AB} \dot x^A\dot x^B-V_Q,
   \eeq
 with the zero-energy constraint
   \beq{5.33n}
     E_Q=\frac12 \bar G_{AB} \dot x^A \dot x^B + V_Q = 0.
   \eeq
 Here
   \beq{5.32n}
     V_Q= \frac12\sum_{I \in \Omega_{e}}
     {\cal E}(I) Q^2(I) \exp[2U^I(x)].
   \eeq

 In section 5 we will examine explicit solutions of the field
 equations that result from (\ref{5.31n}) - (\ref{5.32n}).

    \subsection{Constraints}

 Due to diagonality of the Ricci-tensor for the metric
 (\ref{2.11g}) the non-diagonal part of the Einstein equations
 (\ref{2.4i}) reads as follows
   \beq{2.4ij}
     T_{i j} = 0, \qquad i \neq j.
   \eeq
 This leads to constraints among the charge densities $Q(I)$. First,
 the non-diagonal components of stress-energy tensor are
 proportional to
    \beq{2.4e}
     e^{ 2 \lambda \varphi} F_{i M_2 \dots M_{q}} F_{j}^{~ M_2 \dots M_q},
    \eeq
 with $i \neq j$. From (\ref{2.27n}),  (\ref{5.29na})
 and  (\ref{2.u}) we obtain for the $(p+2)$-form
    \beq{2.27nn}
    F=   \frac{1}{(p+1)!} Q_{i_0 \dots i_p}
       \exp(2 \phi^{i_0} + \dots + 2 \phi^{i_p} - 2 \lambda \varphi)
       du \wedge d y^{i_0} \wedge \dots \wedge d y^{i_p}
   \eeq
 Inserting this in (\ref{2.4e}) we are led to the following
 constraint equations on charge densities \cite{IMC}
    \beq{2.4c}
     C_{ij} \equiv  \sum_{i_1, \dots, i_p =1}^{n}
     Q_{i i_1 \dots i_p} Q_{j i_1 \dots i_p} \eps_{i_1} \e^{2 \phi^{i_1}}
     \dots   \eps_{i_p} \e^{2 \phi^{i_p}} = 0,
    \eeq
 where $i \neq j$; $i, j =1, \dots,n$. $T_{i j}$ is proportional to
 $\exp(- 2 \lambda \varphi - 2 \gamma + 2 \phi^i + 2 \phi^j) C_{i j}$
 for $i \neq j$.

 Here $p = q - 2$  and $Q_{i_0 i_1 .. i_p}$ are components of the
 antisymmetric form of rank $p+1 = q-1$ and
    \beq{2.4d}
     Q_{i_0 i_1 \dots i_p} =  Q(\{ i_0, i_1, \dots , i_p \})
    \eeq
 for $i_0 < i_1 < \dots < i_p$ and $\{ i_0, i_1, \dots , i_p \} \in
 \Omega_e$. The number of constraints in (\ref{2.4c}) is $n(n -
 1)/2$. In the next section we will show that these constraints
 can  be satisfied when the dimension of
 space-time takes certain odd values.

   \section{Solution to constraints on charge densities in various
            dimensions}

 The  constraints (\ref{2.4c}) can be rewritten as
    \beq{3.c}
   \bar{C}_i^j =  \sum_{i_1, \dots, i_p =1}^{n}
     \bar{Q}_{i i_1 \dots i_p} \bar{Q}^{j i_1 \dots i_p} = 0,
    \eeq
 $i \neq j$; $i, j =1, \dots , n$. The charge densities have been
 redefined via
    \beq{3.4d}
     \bar{Q}_{i_0 i_1 \dots  i_p} =  Q_{i_0 i_1 \dots i_p}
     \prod_{k = 0}^{p} \exp(\phi^{i_k})
    \eeq
 and the indices were lifted by the flat metric
    \beq{3.e}
    \eta = \eps _1 dy^1 \otimes dy^1 + \dots + \eps_n dy^n \otimes
    dy^n = \eta _{ab} dy^a \otimes dy^b,
    \eeq
 where $(\eta _{ab}) = (\eta ^{ab}) =
  {\rm diag}(\eps _1 , \dots , \eps _n)$,
 {\it i.e.} $\bar{Q}^{i_0  \dots i_p} = \eta^{i_0 i_0 '} \dots
 \eta^{i_p i_p '} \bar{Q}_{i_0' \dots i_p'}$. ( Here $\bar{C}_i^j =
 C_i^j \exp(\phi^{i} + \phi^{j})$ with $C_i^j = C_{ik} \eta^{kj}$.)

 These $\bar{Q}_{i_0 i_1 \dots  i_p}$ can be viewed as ``running''
 charge densities with the functional dependence coming from
 $\phi^{i_k} (u)$. The charge densities will vary with time or
 spatially depending on whether $u$ is a time-like or space-like
 coordinate respectively.

    \subsection{Maximal  configurations for dimensions $D = 4m + 1$ }

 {\bf D=5 case:} To help illustrate the preceding general analysis
 of the constraints in (\ref{3.c}) we consider the explicit example
 $D=5$, $n = 4$ and $\eps _1 = \dots = \eps _4 =1$. The constraints
 of eqs. (\ref{3.c}) read
     \bear{a1}
     \bar{Q}_{13} \bar{Q}_{23} + \bar{Q}_{14} \bar{Q}_{24} = 0, \\
      \label{a2}
     \bar{Q}_{12} \bar{Q}_{23} - \bar{Q}_{14} \bar{Q}_{34} = 0, \\
      \label{a3}
     \bar{Q}_{12} \bar{Q}_{24} + \bar{Q}_{13} \bar{Q}_{34} = 0, \\
      \label{a4}
     \bar{Q}_{12} \bar{Q}_{13} + \bar{Q}_{24} \bar{Q}_{34} = 0, \\
      \label{a5}
     \bar{Q}_{12} \bar{Q}_{14} - \bar{Q}_{23} \bar{Q}_{34} = 0, \\
      \label{a6}
     \bar{Q}_{13} \bar{Q}_{14} + \bar{Q}_{23} \bar{Q}_{24} = 0.
     \ear
 It is not difficult to verify that the only non-zero solutions  to
 eqs. (\ref{a1})-(\ref{a6}) are
     \bear{b1}
    \bar{Q}_{12} = \mp \bar{Q}_{34}, \\
      \label{b2}
     \bar{Q}_{13} = \pm \bar{Q}_{24}, \\
      \label{b3}
     \bar{Q}_{14} = \mp \bar{Q}_{23}.
     \ear
 This may be obtained by  considering the following three  pairs of
 equations:
      (i) (\ref{a1}) and (\ref{a6});
      (ii) (\ref{a2}) and (\ref{a5});
      (iii) (\ref{a3}) and (\ref{a4}).
 The solution (\ref{b1})-(\ref{b3}) may be written in a compact
 form as
    \beq{3.ea}
    \bar{Q}_{i_0 i_1} =
     \pm \frac{1}{2} \eps _{i_0 i_1 j_0 j_1}
     {\bar Q}^{j_0 j_1} = \pm (* \bar{Q})_{i_0 i_1},
     \eeq
 where $*=*[\eta ]$ is the  Hodge operator with respect to $\eta$.
 That means that any self-dual or anti-self-dual 2-form is the
 solution to a set of quadratic equations (\ref{a1})-(\ref{a6}).

 {\bf D= 4m +1 case:}  We now look at the general case. Based on
 the $D=5$ case we will take the  ``running'' charge density form
 $\bar{Q}_{i_0  \dots i_p}$ as self-dual or anti-self-dual in order
 to satisfy the constraint equations (\ref{3.c}). This
 form can only be self-dual or anti-self-dual when the
 number of the non-distinguished coordinates is twice the rank
 of the form:
     \beq{3.ed}
       n=2(p+1).
     \eeq
 Thus, we restrict ourselves to the case when
     \beq{3.eab}
    \bar{Q}_{i_0  \dots i_p} =
     \pm \frac{1}{(p+1)!} \eps _{i_0  \dots i_p j_0 \dots j_p}
     {\bar Q}^{j_0 \dots j_p} = \pm (* \bar{Q})_{i_0  \dots i_p}.
     \eeq
 Here the symbol $*=*[\eta ]$ is the Hodge operator with respect to
 $\eta$. Squaring the Hodge operator gives
     \beq{3.eb}
     (*)^2 = {\cal E} (-1)^{(p+1)^2} \bf{1},
     \eeq
 where
     \beq{3.ee}
     {\cal E} = \eps _1 \dots \eps_n
     \eeq
 equals $\pm 1$ depending on if there are an even or odd number of
 time-like coordinates.

 It can be easily verified that the set of linear equations
 (\ref{3.eab})  has a non-zero solution  if and only if
     \beq{3.eb1}
     {\cal E} (-1)^{(p+1)^2} = 1.
     \eeq
 The dimension of the space of solutions is
 $\frac{1}{2}C_{2(p+1)}^{p +1}$. The factor of $\frac{1}{2}$ comes from
 (anti-)self-duality condition.

 We  will now demonstrate how having self dual or anti-self dual
 charge density form results in the constraints (\ref{2.4c}) being
 satisfied. First, consider
    \bear{3.f}
     {\bar C}_i ^{~j} &=& \sum_{i_1, \dots, i_p =1}^{n}
     {\bar Q}_{i i_1 \dots i_p} {\bar Q}^{j i_1 \dots i_p} =
     \nonumber \\
     &=& \sum_{i_1, \dots, i_p =1}^{n}  \sum_{j_0, \dots, j_p =1}^{n}
     \pm \frac{1}{(p+1)!} \eps _{i i_1 \dots i_p j_0 \dots j_p}
     {\bar Q}^{j_0 \dots j_p}  {\bar Q}^{j i_1 \dots i_p} ,
    \ear
  where $i \ne j$, and we have used the requirement of self duality
  or anti-self duality for the charge density. This can be further
  rewritten as
     \bear{3.g}
   {\bar C}_i ^{~j} &=&
   \sum_{i_1, \dots, i_p =1}^{n}  \sum_{j_1, \dots, j_p =1}^{n} \pm
   \frac{1}{p!}\eps _{i i_1 \dots i_p j j_1 \dots j_p}
   {\bar Q}^{j j_1 \dots j_p}  {\bar Q}^{j i_1 \dots i_p}
   \nonumber \\
   &=& \sum_{i_1, \dots, i_p =1}^{n}  \sum_{j_1, \dots, j_p =1}^{n}
   \pm \frac{(-1)^p}{p!}\eps _{i j_1 \dots j_p  j i_1 \dots i_p}
   {\bar Q}^{j i_1 \dots i_p} {\bar Q}^{j j_1 \dots j_p}
   \nonumber \\
   &=& (-1)^p {\bar C}_i ^{~j}.
      \ear
  Note that $j$ is not summed over in the two sums above.
  In going from the second line of (\ref{3.f}) to the first
  line of  (\ref{3.g}) we have carried out $p + 1$ identical
  sums with:  $j_0 = j$,  $j_1 = j$, ..., $j_p = j$, respectively.

  From
  (\ref{3.g}) one finds that the constraints in (\ref{3.c}) are
  satisfied automatically for odd $p = 2m - 1$ since
    $${\bar C}_i ^{~j} = -{\bar C}_i ^{~j} \Rightarrow
                            {\bar C}_i ^{~j}=0 ~.$$

  From   (\ref{3.eb1}) we get for odd $p$
    \beq{3.e1}
    {\cal E} = 1,
    \eeq
 i.e. the metric (\ref{3.e}) is either Euclidean
 or has an even number of time-like directions.

 The relationship between $n$ and $p$ is $n=2(p+1)$.  Thus, we find
 non-trivial solutions to the constraints (\ref{3.c}) when the
 total spacetime dimension is
      \beq{3.h}
         D= n+1 = 4m + 1= 5, 9, 13, \dots
      \eeq
 and the signature parameter ${\cal E}$ is positive.

    \subsection{Absence of maximal  configurations for $p =1$
             and even $D$ }

 In this subsection we show that for the even dimensional case ($D
 = 2k$) with 3-form  ($p=1$) there are no solutions with
 maximal number of electric $S1$-branes.

 Here we put  $\eps _1 = \dots = \eps _n =1$.
 Equations  (\ref{3.c}) imply  in this case
       \beq{6.c}
       \sum_{i_1 =1}^{n}
         \bar{Q}_{i i_1} \bar{Q}_{j i_1 } = \delta_{ij} P_i
         \equiv P_{ij}.
       \eeq
 The indices are not summed in the second term.
 Now we assume that all $\bar{Q}_{i j} \neq 0$, $i \neq j$,
 {\it i.e.} a composite 1-brane configuration with maximal number
 of electric branes (``strings'') is considered,
 and show that this leads to an  inconsistency.
 The constants
 $P_i$, which are the values of an $n \times n$ diagonal matrix,
 satisfy $P_i > 0$, since when $i=j$ the first equation is just a
 sum of squares. The exact values of $P_i$'s will not be needed
 here.
 In matrix notation  (\ref{6.c}) reads
       \beq{6.d}
         - \bar{Q}^2 = P,
       \eeq
 where we have used the antisymmetry relation $\bar{Q}_{i j} = -
 \bar{Q}_{ji}$. Calculation of the determinants of the matrices in
 the previous relation leads to
      \beq{6.e}
         (-1)^n ({\rm det} \bar{Q})^2 = {\rm det} P > 0.
       \eeq
 which is not valid for odd $n$. (For odd $n$ ${\rm det} \bar{Q}
 =0$.) Thus, there are no ``maximal'' solutions
 to constraints for even
 dimensions $D$  and $p=1$ in the  model under consideration.

 This implies the absence of ``maximal''
 configurations of composite electric $S1$-branes in
 $10$-dimensional  supergravities and
 low-energy  models of superstring origin
 when only one $3$-form is considered.

 In the next section we examine the cosmological type
 solutions to the field equations for $D=5, 9, 13 ...$ with the
 maximal number of non-zero charge densities  $Q_{j_0 j_1 \dots
 j_p}$ obeying  (\ref{3.eab}).

  \section{Cosmological solutions to the field equations for
            D=5, 9, 13, ...}

  Here we give explicit examples of cosmological
  type solutions   for the dimensions  from (\ref{3.h})
  when the  non-distinguished coordinates are all space-like, {\it i.e.}
    \beq{4.eps}
          \eps _1 = \dots = \eps _n =1.
    \eeq

      First we show that all scale factors are the same
      up to constants:
      \beq{4.ba}
       \phi^i(u) = \phi(u) + c^i.
      \eeq
     From the definition of running constants
    (\ref{3.4d}) and the (anti-) self-duality of the charge
    density form (\ref{3.eab}), it follows
    that
     \beq{4.1}
      \sum_{i \in I} \phi^i = \sum_{j \in \bar{I} } \phi^j
      + {\rm const },
     \eeq
     where  $I$ is an arbitrary subset of $I_0 = \{ 1, \ldots, n \}$
     of length $n/2 = 2m$   and
     \beq{4.2}
       {\bar I} \equiv I_0 \setminus I ,
     \eeq
     is ``dual" set.  For $D = 5$ case  eqs. (\ref{4.1})
     read: $\phi^1 + \phi^2 = \phi^3 + \phi^4   + {\rm const }$
     and two other relations obtained by permutations.

     From relations  (\ref{4.1}) one can
     see that all $\phi^i$ should coincide up to constants
     (for $i \in I$ and  $j \in {\bar I}$ it is sufficient
     to consider another equation with
     $I_1 = (I \setminus \{ i \}) \cup \{ j \})$
     instead of $I$  in (\ref{4.1}) and find
     from both equations that $\phi^i$
     and $\phi^j$ coincide up to a constant).

     Thus, we are led to (\ref{4.ba}).  In what follows
     we put  $c^i = 0$ which may always be done via a proper
     rescaling of $y$-coordinates. This also implies that non-running
     charge density form  $Q_{i_0  \dots i_p}$
     is self-dual or anti-self-dual in a flat
     Euclidean space $\R^n$, i.e.
     \beq{3.eaa}
      Q_{i_0  \dots i_p} =
     \pm \frac{1}{(p+1)!} \eps _{i_0  \dots i_p j_0 \dots j_p}
      Q^{j_0 \dots j_p} = \pm (* Q)_{i_0  \dots i_p}.
     \eeq

        The Lagrangian and total energy constraint are given by
    \bear {4.d}
     L_Q &=& \frac{1}{2}  G_{ij} \dot \phi^i\dot \phi^j-V_Q + \frac{1}{2}
          \dot \varphi ^2, \\
     E_Q &=& \frac{1}{2}  G_{ij} \dot \phi^i \dot \phi^j + V_Q +
     \frac{1}{2}\dot \varphi ^2= 0,
     \ear
     with the potential being
     \beq{4.e}
     V_Q = \frac{1}{2} \sum _I Q^2 (I) \exp \left(2 \sum _{k \in I} \phi
     ^k - 2 \lambda \varphi \right),
     \eeq
   see (\ref{5.31n})-(\ref{5.32n}).

  The field equations for $\phi$ and $\varphi$ from $L_Q$ are
     \bear{4.f}
     \sum _{j= 1} ^n G_{ij} \ddot \phi ^j +
     \sum _I Q^2 (I) \delta ^i _I \exp
     \left( 2 \sum _{k \in I} \phi ^k -2 \lambda \varphi \right) =
     0, \\
     \label{4.fa}
     \ddot \varphi + \sum _I Q^2 (I) (-\lambda) \exp
     \left( 2 \sum _{k \in I} \phi ^k -2 \lambda \varphi \right) =
     0.
     \ear
   Since all $\phi ^i$ satisfy $\phi ^i = \phi$
   one finds $\sum _{j=1} ^n G_{ij} \ddot \phi ^j= \sum _{j=1} ^n
   (\delta _{ij} -1) \ddot \phi ^j = (1-n) \ddot \phi$.
   Next, defining
    \beq{4.q1}
        Q^2  \equiv \sum _I Q^2 (I)  \neq 0
    \eeq
   and noting that
    \beq{4.q2}
     \sum _I Q^2 (I) \delta _I^i =   \frac{1}{2} Q^2
    \eeq
   for any $i = 1, ... ,n$,
   one finds that  the field equations
   (\ref{4.f}) and (\ref{4.fa}) become
     \bear{4.g}
     \ddot \phi &=& \frac{1}{2(n-1)} Q^2
           \exp (n \phi - 2 \lambda \varphi), \\
     \label{4.ga}
     \ddot \varphi &=& \lambda Q^2 \exp (n \phi - 2 \lambda \varphi) .
     \ear
     Finally, since the exponents in (\ref{4.g}) and (\ref{4.ga})
     are the same these two equations can be combined into one as
     \beq{4.h}
     \ddot f = -2 A e^{2 f},
     \eeq
     with the definitions
     \bear{4.ha}
     f &\equiv& \frac{n}{2} \phi - \lambda \varphi, \\
     A &\equiv& \frac{Q^2}{2} K, \qquad
           K \equiv  \lambda ^2 - \frac{n}{4(n-1)} ,
     \label{4.hb}
      \ear
     assumed.

     The first integral of (\ref{4.h}) is
     \beq{4.hc}
       \frac{1}{2} \dot f^2 + A e^{2 f} = \frac{1}{2} C,
     \eeq
     where $C$ is an integration constant.

     Let $K \neq 0$, or
     \beq{lambda}
     \lambda ^2 \neq  \frac{n}{4(n-1)} \equiv \lambda^2_0.
     \eeq

     Equation (\ref{4.h}) has  several solutions:
     \beq{4.i}
        f = - \ln \left[ z |2 A|^{1/2} \right]
     \eeq
     with
     \bear{4.ja}
     z &=& \frac{1}{\sqrt{C}} \sinh \left[ (u-u_0) \sqrt{C} \right],
           \qquad A<0 , ~C>0; \\
                          \label{4.jb}
     &=& \frac{1}{\sqrt{-C}} \sin \left[ (u-u_0) \sqrt{-C} \right],
           \qquad A<0 , ~C<0; \\
                           \label{4.jc}
     &=& u-u_0, \qquad \qquad \qquad \qquad \qquad A<0, ~C=0; \\
                            \label{4.jd}
      &=& \frac{1}{\sqrt{C}} \cosh \left[ (u-u_0) \sqrt{C} \right],
     \qquad A>0 , ~C>0.
     \ear

     One can relate the solutions given in (\ref{4.i}) and
     (\ref{4.ja}) - (\ref{4.jd})
     to $\phi $ and $\varphi$ by using (\ref{4.g}) to construct the
     following relationship
     \beq{4.l}
     \ddot \varphi = 2 (n-1) \lambda \ddot \phi,
     \eeq
     which has the solution
     \beq{4.m}
      \varphi = 2 (n- 1) \lambda \phi + C_2 u + C_1,
     \eeq
     where $C_2 , C_1$ are integration constants. Combining
     (\ref{4.m}) with (\ref{4.ha}) gives
     \beq{4.n}
     \phi = \frac{1}{2 (1- n) K} \left[ \lambda (C_2 u + C_1) +
             f(u) \right], \quad
     \varphi = \frac{n}{4 (1- n) K} \left( C_2 u + C_1 \right)
               - \frac{\lambda f(u)}{K}.
       \eeq

     Applying this to the zero energy constraint
     \beq{4.o}
     E_Q = \frac{1}{2} n (1-n) \dot \phi ^2 +
     \frac{1}{2} \dot \varphi ^2 +
     \frac{1}{2} Q^2 e ^{2 f(u)} = 0
     \eeq
     and using  (\ref{4.hc}) we get
     \beq{4.o1}
      E_Q = \frac{C}{K} -
            \frac{n (C_2)^2}{4K (n-1)} =  0,
     \eeq
     or equivalently,
     \beq{4.o2}
     C =  \frac{n }{4(n-1)} (C_2)^2 \geq 0.
     \eeq
     This tells us that only the three cases
     (\ref{4.ja}), (\ref{4.jc}) and  (\ref{4.jd}) occur
     when real solutions are considered.
     The solution (\ref{4.jb}) will be considered in a
     possible future work with a
     pure imaginary scalar field
     and $\lambda$ (this is equivalent to a ``phantom''
     field that may support the so-called bouncing solution).

     Collecting these results together the solutions for the
     metric, scalar field and $(p + 2)$-form are
     \bear{4.pa}
     ds^2 &=& we^{2n \phi(u)} du^2 + e^{2 \phi(u)}
     \sum _{i=1}^n (dy^i)^2 \\
     \label{4.pb}
     \varphi &=& \frac{n}{4(1-n)K} \left( C_2 u + C_1 \right)
               - \frac{\lambda f(u)}{K}, \\
     \label{4.pc}
     F &=& e^{2 f(u)} du \wedge Q,
     \qquad
     Q =  \frac{1}{(p+1)!} Q_{i_0  \dots i_p}
                           dy^{i_0}  \wedge \dots  \wedge dy^{i_p},
     \ear
     with  $\phi (u)$ given by (\ref{4.n}) and the function
     $f(u)$ given by (\ref{4.i}) and (\ref{4.ja}), (\ref{4.jc}),
     (\ref{4.jd}).  Here the charge density form  $Q$
     of rank $n/2 = 2m$  is self-dual or anti-self-dual in a flat
     Euclidean space $\R^n$:  $Q = \pm * Q$,
     the parameters $C_2, C$ obey (\ref{4.o2}) and the
     dilatonic coupling constant $\lambda$ is non-exceptional,
     see (\ref{lambda}).

    \subsection{Special attractor solution for $C = 0$ }

 Here we examine some  properties of the simplest
 cosmological type solution given in (\ref{4.jc}).  For this
 solution $A < 0$ and hence
      \beq{5.l}
        \lambda ^2 < \lambda^2_0.
      \eeq
 Since $C=0$ for (\ref{4.jc}) one has $C_2 =0$ from the condition
 (\ref{4.o2}). Finally, without loss of generality we put $u_0 =
 0$.

 To get a physical understanding of this solution one should change
 the ``time'' coordinate $u$ to the proper time coordinate $\tau$. In order
 to get the correct sign for the proper time  we fix the sign in
 the relationship between $u$ and the proper time as
     \beq{5.t}
     d\tau = -  e^{n \phi (u)} du,
     \eeq
     where
     \beq{5.ph}
     \phi = \frac{1}{2 (n - 1) K} \ln (u |2 {\bar A}|^{1/2}),
     \qquad
     {\bar A} = A e^{- 2 \lambda C_1}.
     \eeq
 Integrating  (\ref{5.t}) and taking a suitable choice of reference
 point we get
     \beq{5.tau}
     |\alpha | |2 {\bar A}|^{1/2} \tau =
     (u |2 {\bar A}|^{1/2})^{\alpha},
     \eeq
     where $u >0$ and
     \beq{5.al}
     \alpha = \frac{\lambda^2 + \lambda^2_0}{\lambda^2 - \lambda^2_0}
      < 0.
     \eeq
 Since $\alpha <0$, $\tau = \tau (u)$ is monotonically decreasing
 from infinity when $u=0$ to zero when $u=\infty$.

 The metric  (\ref{4.pa}) now reads
     \beq{5.pa}
       ds^2 = w d \tau^2 + B \tau^{2 \nu} \sum _{i=1}^n (dy^i)^2,
     \eeq
 where $\tau > 0$ and
     \beq{5.1a}
       \nu = \frac{2}{n + 4 \lambda^2 (n-1)}, \qquad
        B = (|\alpha| |2 {\bar A}|^{1/2})^{2 \nu}.
     \eeq

 By putting $\lambda =0$ and $w = -1$ in the above
 solution we get a cosmological power-law expansion with a power
 $\nu = 2/n$ that is the same as in the case of $D=1+n$ dust matter with a
 zero pressure (see for example \cite{Lor1,BO,IM2}). This
 is not surprising since it can be argued that the collection of
 branes with charge densities obeying (anti)-self-duality
 condition (\ref{3.eaa}) behaves as a dust matter. Indeed,
 we know that in
 the absence of a scalar field the solution with a single brane
 described by a set $I \subset \{1,...,n \} $ is equivalent to an
 anisotropic fluid with equations of state $p_i = - \rho$ for $i
 \in I$ and $p_i =  \rho$ otherwise, $i = 1,\dots,n$ \cite{IMtop}.
 Here $p_i$ is a pressure in the $i$-th direction and
 $\rho$ is the energy density.
 Thus, our collection of branes is equivalent to a multicomponent fluid
 \cite{IM3}
 with coinciding densities, since all $Q^2(I)$ are equal and
 $\phi^i = \phi$.
 For any $i$ the collection of branes can be split
 into pairs with sets $I, {\bar I}$ such that $i \in I$. The first
 brane gives pressure $p_i^I = - \rho$ and the second one gives
 $p_i^{{\bar I}} =  \rho$ (energy densities are the same ). Hence
 the total pressure is zero for the pair and, thus, for the whole
 collection of  branes.

 Finally, we note that the solution (\ref{5.pa}) is attractor
 solution in the limit  $\tau \to + \infty$,
 or $u \to + 0$,  for the  solutions
 with $A < 0$  given by (\ref{4.ja}). This follows
 just from relation $\sinh u  \sim u$ for small $u$.

    \subsection{Kasner-like behavior for $\tau \to + 0$ }

 We now consider $u \to + \infty$ asymptotical behavior of solutions
 with i) $\sinh$- and ii) $\cosh$- functions corresponding to
 (\ref{4.ja}) and (\ref{4.jd}), respectively. In both cases we find
 that the asymptotic behavior is Kasner like {\it i.e.} the metric
 and scalar field take the form
     \begin{equation}
         ds^2_{as} = w d\tau^2 +
     \sum_{i=1}^{n}\tau^{2 \alpha ^i} A_i (dy^i)^2,
       \qquad
       \varphi_{as} =  \alpha_{\varphi} \ln \tau + \varphi_0,
 \end{equation}
 where $A_i > 0$,  $\varphi_0$ are constants. The Kasner
 parameters obey
 \begin{equation} \label{5.4}
 \sum _{i=1}^n \alpha ^i = \sum _{i=1}^n
 (\alpha ^i) ^2 + (\alpha _{\varphi}) ^2 = 1.
 \end{equation}

 In the first case i) with $\sinh$-dependence and $A < 0$ the
 proper time $\tau$ is decreasing when $u \to + \infty$. From eqs.
 (\ref{4.i}), (\ref{4.ja}),  (\ref{4.jd}), (\ref{4.n}) and  (\ref{5.t})
 we find the  following asymptotic behavior as $u \to + \infty$
     \bear{5.3}
       &&\phi \sim  - b u +  {\rm const}, \qquad
         b =  \frac{\lambda C_2 - \sqrt{C}}{2(n -1) K} > 0,
                     \\   \label{5.3a}
       &&\varphi \sim ( - \lambda_0^2 C_2 +
         \lambda \sqrt{C}) K^{-1} u +  {\rm const},
                     \\   \label{5.3b}
       && \tau \sim {\rm const} \exp( - nb u).
    \ear
 Using these asymptotic relations and writing everything in terms of proper
 time one find that the metric and scalar field take the following
 asymptotic forms
     \bear{5.2}
       &&ds^2_{as} =
       w d\tau^2 + \tau^{2/n} A_0 \sum_{i=1}^{n}(dy^i)^2,
 \qquad
       \varphi_{as} =  \alpha_{\varphi} \ln \tau + \varphi_0,
                     \\   \label{5.2a}
       &&\alpha_{\varphi} ^2 = 1 - \frac{1}{n},
    \ear
 as $\tau \to + 0$. Here  $A_0 > 0$,  $\varphi_0$ are constants. The
 relationship for $\alpha _{\varphi} ^2$ comes from
 (\ref{5.3a}),  (\ref{5.3b}) and agrees with
 (\ref{5.4}) and   $\alpha ^i =1/n$.
 Using (\ref{5.3})-(\ref{5.3b}) one can obtain
 the following relationship: ${\rm sign} [\alpha_{\varphi}] = - {\rm
 sign} [C_{2}]$.

 In the second case ii) with $\cosh$-dependence and $A > 0$ the
 proper time $\tau$ decreases as $u \to +\infty$ for $\lambda C_2 >
 0$  and increases for $\lambda C_2 < 0$. In this case we also get
 an asymptotical Kasner type relations (\ref{5.2})-(\ref{5.2a}) in
 the limit  $\tau \to + 0$ with ${\rm sign} [\alpha_{\varphi}] = -
 {\rm sign}[ \lambda]$. In both $\sinh-$ and $\cosh-$ cases the
 Kasner sets $\alpha = (\alpha^i = 1/n, \alpha_{\varphi})$ obey the
 inequalities
     \beq{5.u}
        U^I(\alpha)= - \lambda \alpha_{\varphi} +
                               \sum_{i \in I} \alpha^i
        = - \lambda \alpha_{\varphi} +  \frac{1}{2} > 0
     \eeq
 for all brane sets $I$. This is in agreement with a general
 prescription of the billiard representation  from \cite{IMb1}.
 In the case ii) we
 obtain the Kasner type behavior (\ref{5.2})-(\ref{5.2a}) in the
 limit $\tau \to + \infty$ with ${\rm sign} [\alpha_{\varphi}] = {\rm
 sign} [\lambda]$.
  In this case
     \beq{5.v}
        U^I(\alpha) <   0
     \eeq
  for all $I$.  In case ii) the scalar field has a bouncing
  behavior in the interval $\tau \in (0, + \infty)$.

    \section{Conclusions}

 In this article we have examined a system where $D=n+1$
 dimensional gravity was coupled to a scalar field plus a $p+2$
 rank form field. The ansatz employed here was to take the metric
 as diagonal, and the rank $p+2$ form field to have a composite
 electric $Sp$-brane form. All ansatz functions depended only on
 the one distinguished coordinate, $u$. Under these conditions the
  initial model could be reduced to an effective 1-dimensional
 $\sigma$-model  which greatly simplified the study of
 the system. The diagonal character of our metric ansatz resulted
 in there being constraint equations among the  charge
 densities of branes associated with the $p+2$ form field. By examining
 these constraint equations we showed that  for certain odd values of the
 spacetime dimension, given by $D=4m+1 =5, 9, 13...$, the system
 allowed the maximal number of the charged  branes.
 We also proved the absence of such maximal configurations
 for $p = 1$ and even $D$.

 For these special odd dimensions and
 non-exceptional dilatonic coupling
 ($\lambda^2 \neq  \frac{n}{4(n-1)}$)
 we wrote down exact  solutions given in equations
 (\ref{4.pa})-(\ref{4.pc}).
 We examined   the simplest of these solutions given by
 (\ref{4.jc}). On converting the distinguished coordinate $u$ to
 the proper time $\tau$, we found this solution corresponded to a
 cosmological power-law expansion. By taking the limit of vanishing
 dilatonic coupling, $\lambda =0$, we showed that this solution
 reduced to a cosmological model with dust matter.
 On physical grounds this is exactly what one would expect in this
 limit. We also investigated an asymptotical Kasner type behavior
 of the solutions for small ($\tau \to + 0$) and large ($\tau \to +
 \infty$) values of proper time.

 Directions for possible future work include: (i) the investigation
 of the properties of the solution in (\ref{4.jb}) using pure
 imaginary scalar field and $\lambda$ (with a hope to investigate
 the possibility of bouncing behavior); (ii) the investigation of
 static, non-cosmological solutions, when $w=+1$ and the
 distinguished coordinate $u$ is spacelike; (iii) examining the
 exceptional  case, when $K=0$ ({\it i.e.} $\lambda^2 = \frac{n}{4(n-1)}$)
 in (\ref{4.hb}).

 Using the results obtained in this paper and in \cite{IMb1} we
 will present in  future work \cite{IS}
 some examples of breaking of the never-ending
 oscillating behavior near the singularity by constraints.
 Physically this may indicate that the form of the ansatz considered
 here (diagonal metric, composite electric ansatz for the form
 field, {\it etc.}) may prevent the occurrence of  the  ``chaotic",
 never-ending oscillating behavior (see
 \cite{IMb1,DH,DHN,BKL,IKM} and references, therein).

 It should be noted that recently N.S. Deger has
 obtained in \cite{Deg}  some  $S$-brane solutions
 in $D= 11$ supergravity  with  non-standard intersection
 rules.  These solutions contain ``dangerous'' intersections of branes
 and give certain examples of  solutions
 with charge densities obeying constraint equations. It seems
 that the ``non-standard'' solutions  from \cite{Deg}
 with vanishing Chern-Simons term  may be obtained
 as a special case of the so-called ``block-orthogonal'' solutions
 \cite{Br1,IMJ2,IMJ1}. In these block-orthogonal solutions the branes split
 into several blocks.  The branes belonging to one block behave as one
 ``effective'' brane. The branes belonging to different blocks
 obey the ``standard'' intersection rules \cite{IMC,AR,AEH}.
 In  future work \cite{IS} we will verify whether the solution
 from our paper may be obtained using the
 modified one-block solution from  \cite{IMJ2,IMJ1}.

 \begin{center}
 {\bf Acknowledgments}
 \end{center}

 The work of V.D.I. was supported in part by a DFG grant. The work of
 D.S. was supported by a 2003-2004 Fulbright Fellowship.

   \np


\small
   \begin{thebibliography}{99}

   \bibitem{BGrIM}
   K.A. Bronnikov, M.A. Grebeniuk, V.D. Ivashchuk and V.N. Melnikov,
   {\it Integrable Multidimensional Cosmology
   for Intersecting p-branes}, {\it Grav. Cosmol.},
   {\bf 3}, No 2 (10), 105 (1997).

   \bibitem{GrIM}
   M.A. Grebeniuk, V.D. Ivashchuk and V.N. Melnikov,
   {\it Integrable multidimensional quantum cosmology  for intersecting p-branes},
   {\it Grav. and Cosmol.\/} {\bf 3}, No 3 (11), 243 (1997)
   [arXiv:gr-qc/9708031].

   \bibitem{IMJ}
   V.D. Ivashchuk and V.N. Melnikov,
   {it Multidimensional classical and quantum cosmology
   with intersecting $p$-branes},
   {\it J. Math. Phys.}, {\bf 39}, 2866 (1998)
   [arXiv:hep-th/9708157].

   \bibitem{IK}
   V.D. Ivashchuk and S.-W. Kim, {\it Solutions with intersecting p-branes
   related to Toda chains}, {\it J. Math. Phys.}, {\bf 41} (1)
   444 (2000)
   [arXiv:hep-th/9907019].

   \bibitem{IMtop}
   V.D. Ivashchuk and V.N. Melnikov,  {\it Exact solutions in
   multidimensional gravity with antisymmetric forms}, 
   {\it Class. Quantum Grav.} {\bf 18},
   R82-R157 (2001) [arXiv:hep-th/0110274].

    \bibitem{S1}
    M. Gutperle and A. Strominger,  {\it Spacelike branes}, {\it JHEP}
    {\bf 0204}, 018 (2002)
    [arXiv:hep-th/0202210].

    \bibitem{S2}
    C.M. Chen, D.M. Gal'tsov and M. Gutperle,  {\it S-brane
    solutions in supergravity theories},
    {\it Phys. Rev.}  {\bf D 66}, 024043 (2002)
    [arXiv:hep-th/0204071].

    \bibitem{Isbr}
    V.D. Ivashchuk, {\it Composite S-brane solutions
    related to Toda-type systems}, {\it Class. Quantum
    Grav.} {\bf 20}, 261 (2003)
    [arXiv:hep-th/0208101].

    \bibitem{Ohta}
    N. Ohta,  {\it Intersection rules for S-branes},
    {\it Phys. Lett.}  {\bf B 558}, 213 (2003)
    [arXiv:hep-th/0301095].

    \bibitem{Iohta}
    V.D. Ivashchuk,
    {\it On composite S-brane solutions with orthogonal
    intersection rules} [arXiv:hep-th/0309027].

    \bibitem{Ierice}
    V.D. Ivashchuk, {\it On exact solutions in
    multidimensional gravity with antisymmetric forms},
    In: Proceedings of the 18th Course of the School on
    Cosmology and Gravitation: The Gravitational Constant.
    {\it Generalized Gravitational Theories and Experiments} (30 April-10
    May 2003, Erice).  Ed. by G.T. Gillies, V.N. Melnikov and V. de
    Sabbata, (Kluwer Academic Publishers, Dordrecht, 2004), pp. 39-64;
    [arXiv:gr-qc/0310114].

   \bibitem{IMC}
   V.D. Ivashchuk and V.N. Melnikov,
   {\it Sigma-model for the generalized  composite p-branes},
   {\it Class. Quantum Grav.} {\bf 14}, 3001 (1997);
   Corrigenda {\it ibid.} {\bf 15 } (12), 3941 (1998)
   [arXiv:hep-th/9705036].

  \bibitem{IMb1}
  V.D. Ivashchuk and V.N. Melnikov, {\it Billiard representation for
  multidimensional cosmology with intersecting p-branes near the
  singularity}.  {\it J. Math. Phys.},  {\bf 41}, No 8, 6341 (2000)
  [arXiv:hep-th/9904077].


   \bibitem{DH}
   T. Damour and  M. Henneaux,
   {\it Chaos in superstring cosmology},
   {\it Phys. Rev. Lett.} {\bf 85}, 920 (2000)
   [arXiv:hep-th/0003139].

   \bibitem{DHN}
   T. Damour, M. Henneaux and H. Nicolai,
   {\it Cosmological billiards, topical review}, {\it Class. Quantum
   Grav.} {\bf 20}, R145-R200 (2003)
   [arXiv:hep-th/0212256].


   \bibitem{IM2}
    V.D. Ivashchuk and V.N. Melnikov, {\it Perfect-fluid type  solution
    in multidimensional cosmology}, {\it Phys. Lett. } {\bf A 135}, 9,
    465 (1989).

   \bibitem{IMZ}
   V.D. Ivashchuk, V.N. Melnikov and A.I. Zhuk,
   {\it On Wheeler-DeWitt  equation in multidimensional cosmology},
   {\it Nuovo Cimento } {\bf B 104}, 575  (1989).

   \bibitem{Lor1}
   D. Lorentz-Petzold, {\it Higher-dimensional cosmologies}, {\it Phys.
   Lett. } {\bf B 148}, 43 (1984).

   \bibitem{BO}
   R. Bergamini and C.A. Orzalesi, {\it Towards a cosmology for
   multidimensional unified theories}, {\it Phys. Lett. } {\bf B
   135}, 38 (1984).


   \bibitem{IM3}
    K.A. Bronnikov and V.N. Melnikov,  {\it The Birkhoff theorem in
    multidimensional gravity}, {\it Gen. Rel. Grav.} {\bf 27}, (1995) 465
    [arXiv:gr-qc/9403063].


  \bibitem{BKL}
  V.A. Belinskii, E.M. Lifshitz  and I.M. Khalatnikov,
  {\it Oscillatory approach to a singular point in the relativistic cosmology},
  {\it Adv. Phys.}  {\bf 19} (1970) 525; {\it General solution of the Einstein 
  with a time singularity}, {\it Ann. Phys. (NY)} {\bf 31} (1982) 639.

  \bibitem{IKM}
   V.D. Ivashchuk, A.A. Kirillov and V.N. Melnikov, {\it Stochastic properties of 
   multidimensional cosmological models near a singular point}, {\it JETP Lett.}
   {\bf 60} (1994) 235 [{\it Pis'ma ZhETF } {\bf 60}, (1994) 225].

   \bibitem{Deg}
   N.S. Deger, {\it Non-standard intersections of S-Branes in D=11
   supergravity} {\it JHEP} {\bf 0304} (2003) 034
   [arXiv:hep-th/0303232].


  \bibitem{Br1}
  K.A. Bronnikov, {\it Block-orthogonal brane systems, black
  holes and wormholes},
  {\it Grav. and Cosmol.} {\bf 4}, No 1 (13), 49 (1998)
  [arXiv:hep-th/9710207].

 \bibitem{IMJ2}
 V.D. Ivashchuk and V.N. Melnikov, {\it Multidimensional cosmological and
 spherically symmetric solutions with intersecting p-branes.}
 In Lecture Notes in Physics, Vol. 537,
 {\it Mathematical and Quantum Aspects of Relativity and Cosmology
 Proceedings of the Second Samos Meeting on Cosmology, Geometry and
 Relativity} held at Pythagoreon, Samos, Greece, 1998,
 eds: S. Cotsakis, G.W. Gibbons., Berlin, Springer, 2000
 [arXiv:gr-qc/9901001].

  \bibitem{IMJ1}
  V.D. Ivashchuk and V.N. Melnikov, {\it Cosmological and spherically
  symmetric solutions with intersecting p-branes}. {\it J. Math. Phys.},
  {\bf 40} (12), 6558 (1999).

  \bibitem{AR}
  I.Ya. Aref'eva and O.A. Rytchkov,
  {\it Incidence matrix description of intersecting p-brane solutions},
  {\it Am. Math, Soc. Transl.} {\bf 201}, 19 (2000) [arXiv:hep-th/9612236];
  I.Ya. Aref'eva, M.G. Ivanov and I.V. Volovich, {\it Non-extremal intersecting p-branes
  in various dimensions}, {\it Phys. Lett.} {\bf B 406}, 44 (1997) [arXiv:hep-th/9702079].

  \bibitem{AEH}
  R. Argurio, F. Englert and L. Hourant,
  {\it Intersection rules for $p$-branes},
  {\it Phys. Lett. } {\bf B 398}, 61 (1997) [arXiv:hep-th/9701042];
  N. Ohta, {\it Intersection rules for non-extreme p-branes}, {\it Phys. Lett.}
  {\bf B 403}, 218 (1997) [arXiv:hep-th/9702164] 

   \bibitem{IS}
   V.D. Ivashchuk and D. Singleton, in preparation

  \end{thebibliography}
  \end{document}